\documentclass[twocolumn,showpacs,preprintnumbers,amsmath,amssymb,prb]{revtex4}

\usepackage{graphicx}
\usepackage{dcolumn}
\usepackage{bm}

\begin{document}


\title{Two-photon up-conversion affected by inter-molecule correlations \\ near metallic nanostructure}

\author{Yoshiki Osaka}
 \email{osaka@pe.osakafu-u.ac.jp}
 \affiliation{Department of Physics and Electronics, Osaka Prefecture University, 1-1 Gakuen-cho, Sakai, Osaka 599-8531, Japan}
\author{Nobuhiko Yokoshi}
 \email{yokoshi@pe.osakafu-u.ac.jp}
 \affiliation{Department of Physics and Electronics, Osaka Prefecture University, 1-1 Gakuen-cho, Sakai, Osaka 599-8531, Japan}
\author{Hajime Ishihara}
\affiliation{Department of Physics and Electronics, Osaka Prefecture University, 1-1 Gakuen-cho, Sakai, Osaka 599-8531, Japan}

\date{\today}

\begin{abstract}
We investigate an efficient two-photon up-conversion process in more than one molecule coupled to an optical antenna. In the previous work [Y. Osaka {\it et al.}, PRL {\bf 112}, 133601 (2014)], we considered the two-photon up-conversion process in a single molecule within one-dimensional input-output theory, and revealed that controlling the antenna-molecule coupling enables the efficient up-conversion with radiative loss in the antenna suppressed. In this work, aiming to propose a way to enhance the total probability of antenna-photon scattering, we extend the model to the case of multiple molecules. In general, the presence of more than one molecule decreases the up-conversion probability because they equally share the energy of the two photons. However, it is shown that we can overcome the difficulty by controlling the inter-molecule coupling. Our result implies that, without increasing the incident photon number (light power), we can enlarge the net probability of the two-photon up-conversion.
\end{abstract}


\maketitle

\section{introduction \label{sec.intro}}
Utilizing absorption saturation of discrete levels in nanomaterials such as molecules and quantum dots is a useful way to enhance optical nonlinearities by few photons~\cite{discrete}.
However, it is necessary to supplement small absorption cross sections of nanomaterials for photons with auxiliary systems because the size of nanomaterials is tiny compared to the spatial extent of photon wave functions. Optical cavity is well-known as an example of the auxiliary system to enhance interactions between photons and nanomaterials~\cite{MicroCavity,cavity}. Another approach is to introduce optical antenna that consists of metallic nanostructures. It produces electric fields localized beyond the diffraction limit near the metal surface by localized surface plasmon resonance~\cite{Muehlschlegel2005,antenna1}. By embedding the nanomaterials near the optical antenna, the nanomaterials strongly interact with the localized fields~\cite{Anger2006,Kinkhabwala2009}. Especially, when metallic nanostructres are set at intervals of a few nanometers, there exists a ``hot spot'' near the gap region where the field intensity is enhanced up to $10^5$-fold~\cite{Hao2004}. In such a hot spot, the conventional selection rules of optical transitions are broken because of high gradient intensity of the strongly localized field~\cite{forbidden,forbidden2,Takase2013}. In addition to the potential as the optical antenna, surface plasmon has a potential to combine an enormous capacity of photonics and a miniaturization of electronics~\cite{PlasmonSubwavelengthOptics,MergingPhotonicsElectronics,PromisePlasmonics}. However, due to the large radiative and nonradiative losses of the plasmon~\cite{ReviewNanoplasmonics,PhysRevLett.84.5644,PhysRevLett.88.077402}, it is difficult to efficiently excite nanomaterials by weak light. If the losses in the antenna are successfully suppressed, we can expect the antenna-assisted system to be applied to key technologies in quantum information, communication and computation such as visible-to-telecom frequency conversion of single photons emitted by a quantum dot~\cite{SingleVTconv} and single-photon switch~\cite{2photonGate,Volz2012}. 

In the previous work, we theoretically demonstrated optical linear responses on a antenna-molecule coupled system and reported that the molecule efficiently absorbs the incident light energy with the loss in the antenna suppressed. This is because the interference in the coupled modes makes the antenna mode latescent~\cite{Ishihara2011,Nakatani2013}. Moreover, we revealed that this phenomenon is significantly beneficial in nonlinear optical responses~\cite{OsakaPRL2014}. As a next step toward applications, we have to explore the way to enhance the total scattering probability between the antenna-molecule coupled system and photons. Because the wavelength of the photon is much larger than the size of the coupled system, the antenna effect is limited. In order to overcome the difficulty, we consider larger antenna-molecule coupled system involving more than one molecule. However, the presence of more than one molecule seems to disturbs nonlinear processes because of the inhibition of photon-photon interaction in individual molecules, i.e., the optical nonlinearity requires incident light with higher intensity.

In this paper, we analyze the up-conversion process by two photons in more than one molecule coupled to an optical antenna in order to investigate the effect of the presence of the multi-molecules. In consequence, we confirm that the up-conversion probability under two-photon irradiation decreases with increasing the number of molecules. However, we also find that the up-conversion efficiently occurs at the optimal inter-molecule coupling. This is because the inter-molecule coupling lifts the degeneracy of the energy levels of the molecular system so that the offending interference due to the presence of more than one molecule is avoided. Recently, we reported that the radiation-induced coupling between molecules nearby a metallic nanostructure can become considerable even when the inter-molecule distance is tens of nanometers~\cite{OsakaAntennPropag2015}. This means that the inter-molecule coupling in large antenna-molecule system is achievable. These results indicate that controlling the inter-molecule coupling provides high-efficient few-photon nonlinear responses even in the multi-molecular system without increasing the photon number. In addition, utilizing photons with quantum correlation for two-photon processes has attracted much attention in terms of not only basic science but also applications~\cite{TPA1,TPA2,TPA3,TPA4,PhysRevLett.103.123602,TPA5}, because entangled photons are key issues in quantum information technology~\cite{NC} and the generation efficiency of them is largely growing recently~\cite{PDC,HPS,ELED,QD&I}. Therefore, we also examine the dependence of nonlinear optics on photon correlations, and find that the correlated photons facilitate the up-conversion process.

The paper is organized as follows. In Sec.~\ref{sec.formalism}, we introduce the theoretical formalism to analyze the two-photon up-conversion process in the system where more than molecule is coupled to a metallic nanoantenna. In Sec.~\ref{sec.results}, we show the numerically calculated up-conversion probabilities, which is enhanced in the suitable conditions, and then discuss the essence of the enhancement. We also investigate the dependence of up-conversion on the correlation between input photons. Our summary and conclusion are provided in Sec.~\ref{sec.conclusion}.

\section{formalism \label{sec.formalism}}
\subsection{Multi-molecular system coupled to an antenna \label{subsec.model}}
\begin{figure}[h]%
\includegraphics[width=80mm]{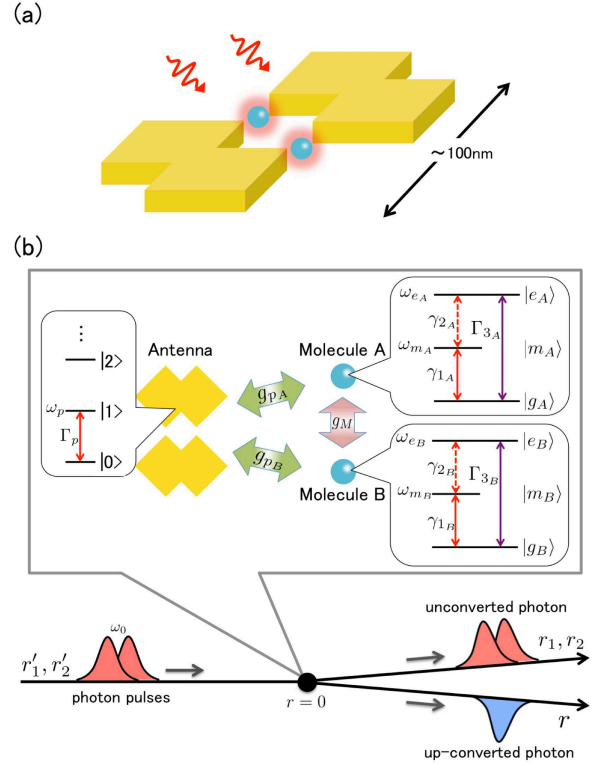}
\caption{(a) Illustration of one example of the optical nanoantenna structure that couples to two molecules. (b) Schematic illustration of the coupled antenna-molecules system. The localized cavity mode with large relaxation constant $\Gamma_p$ excites the four-level molecule. The relaxation constants in the molecule A(B) $\{\gamma_{1_{A(B)}}, \gamma_{2_{A(B)}}, \gamma_{3_{A(B)}}\}$ are set to be much smaller compared with $\Gamma_p$. The constant $g_{p_A,p_B}$ and $g_M$ denote the antenna-molecule and the molecule-molecule coupling, respectively.}
\label{fig.model}
\end{figure}

We consider the up-conversion process by two photons in a multi-molecular system coupled to an optical antenna. Although, in this section, we show the most simple case that the system contains two molecules and an optical antenna, we can also analyze the up-conversion in more than two molecules in the same manner. Therefore, in the following sections, we show the results in the case that the number of molecules is two, three, and four.

In Fig.~\ref{fig.model}(a), we show one example of the system setup. The metallic nanoantenna has two hot spots near which molecules are embedded. Such a nanoantenna is actually fabricated using electron beam lithography and lift-off technique~\cite{Tanaka2015}. In addition, the recent progress of nanofabrication techniques enables us to prepare arrays of metallic nanostructures~\cite{Ueno2008,PhysRevLett.101.087403,Roxworthy2014}. It should be noted that the typical size of the antennas is of the order of 100 nm, which is small compared to a spatial extent of photons. Therefore, even though the antenna includes more than one and separated hot spots, the excitation by a single photon must be regarded as a single surface plasmon. Therefore, we consider a simplified model shown in Fig.~\ref{fig.model}(b) (in the case of two molecules). The antenna mode is modeled by simple boson with the plasmon resonance frequency $\omega_p$ and the large radiative decay rate $\Gamma_p$. The molecule A(B) is considered as a three-level system, in which the states $|m_{A(B)}\rangle$ and $|e_{A(B)}\rangle$ are dipole-allowed from the ground state $|g_{A(B)}\rangle$, and the transition between the state $|e_{A(B)}\rangle$ and $|m_{A(B)}\rangle$ is dipole-forbidden owing to the parity of the wave function. The radiative decay rates of the molecule A(B) are denoted by $\{\gamma_{1_{A(B)}}, \gamma_{2_{A(B)}}, \Gamma_{3_{A(B)}}\}$, where the rates of the forbidden transitions $\gamma_{2_{A(B)}}$ are much smaller than the ones of the allowed transitions $\{\gamma_{1_{A(B)}}, \Gamma_{3_{A(B)}}\}$. We take the molecule-molecule coupling given by the constant $g_M$ in this model because it is known that the coupling between emitters located near surface of a metal or graphene is enhanced through the plasmons~\cite{PhysRevLett.102.077401,PhysRevB.85.155438,OsakaAntennPropag2015}. The constant $g_{p_{A(B)}}$ denotes the coupling between the antenna and the molecule A(B). The total Hamiltonian is 
\begin{align}
{\cal H}=&{\cal H}_{\rm antenna}+{\cal H}_{\rm molecule}+{\cal H}_{\rm photon}
\nonumber\\
&+{\cal H}_{\rm a-p}+{\cal H}_{\rm m-p}+{\cal H}_{\rm a-m}+{\cal H}_{\rm m-m}.
\end{align}
Here the energy of the antenna mode is described as ${\cal H}_{\rm antenna}=\hbar \omega_p p^{\dagger}p$, where the operator $p^{(\dagger)}$ annihilates (creates) an antenna mode.
As for the molecules, the Hamiltonian is ${\cal H}_{\rm molecule}=\hbar \omega_{m_A}\sigma_{Amm}+ \hbar \omega_{e_A}\sigma_{Aee}+\hbar \omega_{m_B}\sigma_{Bmm}+ \hbar \omega_{e_B}\sigma_{Bee}$, in which $\sigma_{A(B)ij} =|i_{A(B)}\rangle \langle j_{A(B)}|$ with  $\{i,j \} = \{ g,m,e \}$ and the resonant energies are measured from the ground state $|g_{A(B)}\rangle$. We employ the one-dimensional model and then the energy of the photon field can be described as ${\cal H}_{\rm photon}=\int dk \hbar ck (\tilde{a}_{k}^{\dagger}\tilde{a}_{k}  + \tilde{b}_{k}^{\dagger}\tilde{b}_{k})$. Here the operator $b_r^{(\dagger)}$ annihilates(creates) an up-converted photon at position $r$,  and  $a_r^{(\dagger)}$ annihilates(creates) an input or unconverted output photon. The tilde on the operators indicates Fourier transformation, e.g., $\tilde{a}_k=\sqrt{1/2\pi} \int  dr  a_r e^{-ikr}$. Input two photons interact with the coupled system at the origin $(r=0)$. The antenna-photon coupling is written within rotating-wave approximation as 
\begin{align}
{\cal H}_{\rm a-p}=i \hbar \sqrt{c\Gamma_p} (  p^{\dagger}a_{r=0} -a_{r=0}^{\dagger} p  ).
\end{align}
In the similar manner, the radiations of the photons from the molecule are 
\begin{align}
{\cal H}_{\rm m-p}=&i \hbar \sqrt{c\gamma_{1_A}}  \sigma_{Amg}a_{r=0}
+i \hbar \sqrt{c\gamma_{2_A}} \sigma_{Aem}a_{r=0}
\nonumber\\&
+i \hbar \sqrt{c\gamma_{1_B}} \sigma_{Bmg}a_{r=0} 
+i \hbar \sqrt{c\gamma_{2_B}} \sigma_{Bem}a_{r=0} 
\nonumber\\&
+i \hbar \sqrt{c\Gamma_{3_A}} \sigma_{Aeg}b_{r=0} 
+i \hbar \sqrt{c\Gamma_{3_B}}  \sigma_{Beg}b_{r=0}
\nonumber\\&
+ \rm{h.c.}.
\end{align}
The interaction of the molecules with the plasmon field is described as 
\begin{align}
{\cal H}_{\rm a-m}=&
\hbar g_{p_A} \left(  \sigma_{Amg}p 
                             +\sigma_{Aem}p 
 \right)
\nonumber\\
&
+\hbar g_{p_B} \left(  \sigma_{Bmg}p 
                             +\sigma_{Bem}p 
\right)
+\rm{h.c.},
\end{align}
where the localized field with spatial gradient is assumed to produce similar intensities of absorptions in both the dipole-allowed and -forbidden transitions~\cite{forbidden,forbidden2,Takase2013}. The molecule-molecule interaction via the radiation field is written as
\begin{align}
{\cal H}_{\rm m-m}=&
\hbar g_M \left(   \sigma_{Amg}\sigma_{Bgm} 
+ \sigma_{Amg}\sigma_{Bme} 
\right.\nonumber\\
&\left. + \sigma_{Aem}\sigma_{Bgm} 
+ \sigma_{Aem}\sigma_{Bme} 
 \right)
+\rm{h.c.}.
\end{align}
It is known that the coupling between emitters located near surface of a metal or a graphene is enhanced through the plasmons~\cite{PhysRevLett.102.077401,PhysRevB.85.155438}. Moreover, the coupling between molecules that are embedded near gaps of metallic nanostructures is comparable in magnitude to plasmon-molecule coupling even when the one molecule is separated from the other molecule by tens of nanometers~\cite{OsakaAntennPropag2015}. Therefore, we can acquire the inter-molecule coupling of the near order of the antenna-molecule coupling. We analyze the up-conversion process by applying the input-output formalism to this model.

\subsection{Up-conversion process \label{subsec.input-output}}
As an initial state, we consider a two-photon state. Then, the initial state vector is written as 
\begin{align}
|\Psi_{\rm in}\rangle
&=\iint dr'_{1}dr'_{2} \frac{f(r'_{1},r'_{2})}{\sqrt{2!}}a_{r'_1}^\dagger a_{r'_2}^\dagger|V \rangle.
\end{align}
On the other hand, the output state is a superposition of a up-converted photon state and two-photon state. Therefore, the output state vector can be written as
\begin{align}
|\Psi_{\rm out}\rangle
&=e^{-i{\cal H}t}|\Psi_{\rm in}\rangle
\nonumber\\
&=\int_{0}^{\infty}  dr ~ h(r;\tau)b_r^\dagger|V \rangle
\nonumber\\
&~~~+\iint_{0}^{\infty}  dr_1 dr_2 ~  \frac{g(r_1,r_2;\tau)}{\sqrt{2!}}a_{r_1}^\dagger a_{r_2}^\dagger|V \rangle.
\end{align}
Here the wave function of the up-converted photon can be written as
\begin{align}
 h(r;\tau)
&=\langle V | b_ r |\Psi_{\rm out}\rangle =\langle V | b_ r (\tau) |\Psi_{\rm in}\rangle \nonumber\\ 
&=\iint dr'_{1}dr'_{2} {\rm G_{uc}}(r,r'_1,r'_2;\tau) f(r'_{1},r'_{2}),
\end{align}
where $b_r(\tau)=e^{i{\cal H}\tau} b_r e^{-i{\cal H}\tau}$ (Heisenberg picture), and ${\rm G_{uc}}(r,r'_1,r'_2;\tau)$ is the propagator for the up-conversion process, which is defined by 
\begin{align}
{\rm G_{uc}}(r,r'_1,r'_2;\tau)
=\frac{\langle V| b_{r} (\tau) a^{\dagger}_{r'_{1}} a^{\dagger}_{r'_{2}}|V \rangle}{\sqrt{2}}.
\label{eq.G_up}
\end{align}
We derive the propagator in order to gain the wave function of the up-converted photon.
In the same manner, the propagator for the two-photon emission process is given as
\begin{align}
{\rm G_{two}}(r_1,r_2,r'_1,r'_2;\tau)
=\frac{\langle V| a_{r_1}(\tau) a_{r_2}(\tau) a^{\dagger}_{r'_1} a^{\dagger}_{r'_2}|V \rangle}{2}.
\end{align}

In calculating the propagators, we employ the method developed in Ref.~\onlinecite{Koshino2009}, in which a coherent state of the photon field is introduced.
According to this method, we define the coherent state as
\begin{align}
|\phi \rangle ={\cal N}\exp(\mu_{1}a_{r'_{1}}^{\dagger}+\mu_{2}a_{r'_{2}}^{\dagger})|V \rangle,
\label{coherent state eq.}
\end{align}
where ${\cal N}$ is normalization factor and $\mu_{1,2}$ are perturbation coefficients. For the above state, one can write down the relation
\begin{align}
a_r| \phi \rangle&= \sum_{j=1,2} \mu_j \delta(r-r_j')| \phi \rangle,
\\
b_r| \phi \rangle&=0.
\end{align}
From the Heisenberg equations for the operators of the photons, output fields are obtained as 
\begin{align}
a_{r}(\tau)=&
a_{r-c\tau}(0)
\nonumber\\
&- \left\{
\sqrt{\frac{\Gamma_p}{c}}  p(\tau-\frac{r}{c})
 + \sqrt{\frac{\gamma_{1_A}}{c}}  \sigma_{Agm}(\tau-\frac{r}{c})
 \right. \nonumber\\&
+ \sqrt{\frac{\gamma_{2_A}}{c}}  \sigma_{Ame}(\tau-\frac{r}{c}) 
+ \sqrt{\frac{\gamma_{1_B}}{c}}  \sigma_{Bgm}(\tau-\frac{r}{c})  
 \nonumber\\&
\left. + \sqrt{\frac{\gamma_{2_B}}{c}}  \sigma_{Bme}(\tau-\frac{r}{c}) 
\right\}
\left\{\theta(\frac{r}{c})-\theta(\frac{r}{c}-\tau)\right\},
\label{a_r(t)}
\end{align}
\begin{align}
b_{r}(\tau)
=&
b_{r-c\tau}(0)
\nonumber\\
&-   \left( \sqrt{\frac{\Gamma_{3_A}}{c}}   \sigma_{Age}(\tau-\frac{r}{c}) + \sqrt{\frac{\Gamma_{3_B}}{c}}  \sigma_{Bge}(\tau-\frac{r}{c}) \right) 
 \nonumber\\&\times
\left\{\theta(\frac{r}{c})-\theta(\frac{r}{c}-\tau)\right\},
\label{b_r(t)}
\end{align}
where $\theta(\tau)$ is Heaviside step function.
From Eqs.~(\ref{eq.G_up}) and (\ref{b_r(t)}), we find the propagator for the up-conversion process to be
\begin{align}
&{\rm G_{uc}}(r'_1,r'_2,r;\tau)
\nonumber\\
&=\frac{\langle  b_{r} (\tau) \rangle^{\mu_{1}\mu_{2}}}{\sqrt{2}}\nonumber\\
&=-\sqrt{\frac{\Gamma_{3_A}}{2c}}   \langle  \sigma_{Age}(t-\frac{r}{c}) \rangle^{\mu_1\mu_2} \left\{\theta(\frac{r}{c})-\theta(\frac{r}{c}-t)\right\} \nonumber\\
&~~~ -\sqrt{\frac{\Gamma_{3_B}}{2c}}  \langle   \sigma_{Bge}(t-\frac{r}{c}) \rangle^{\mu_1\mu_2} 
\left\{\theta(\frac{r}{c})-\theta(\frac{r}{c}-t)\right\},
\label{eq.uppropagator5}
\end{align}
where $\langle  b_{r} (\tau) \rangle^{\mu_{1}\mu_{2}}$ means the perturbation component of proportional to $\mu_1\mu_2$ in $\langle b_{r} (\tau) \rangle =\langle \phi| b_{r} (\tau) |\phi \rangle$. Therefore, we can calculate the wave function of the up-converted photon  when we get $\langle   \sigma_{Age}(\tau) \rangle^{\mu_{1}\mu_{2}}$ and $\langle   \sigma_{Bge}(\tau) \rangle^{\mu_{1}\mu_{2}}$.

We can obtain time evolution equations for the operators from Heisenberg equations, i.e., 
\begin{align}
\frac{d}{d\tau}\sigma_{Age}(\tau)
&=
\frac{i}{\hbar}[{\cal H},\sigma_{Age}(\tau)].
\end{align}
From the above equation, the equation for the expectation value of the operator  $\sigma_{Age}(\tau)$, i.e., $\langle \sigma_{Age}(\tau)  \rangle$ is obtained.
Furthermore, the equation of motion for $ \langle \sigma_{Age}(\tau)  \rangle ^{\mu_1\mu_2}$ can be written as 
\begin{align}
&\frac{d}{dt} \langle \sigma_{Age}(\tau) \rangle ^{\mu_1 \mu_2}\nonumber\\
&=-(i\omega_{e_A}+\frac{\Gamma_{3_A}+\gamma_{2_A}}{2}  )  \langle \sigma_{Age}(\tau) \rangle ^{\mu_1 \mu_2}
\nonumber\\&
- \left( i g_{p_A} + \frac{\sqrt{\gamma_{2_A}\Gamma_p}}{2} \right)  \langle \sigma_{Agm}(\tau)p(\tau)  \rangle ^{\mu_1 \mu_2} 
\nonumber\\&
+ \sqrt{c\gamma_{2_A}} \langle \sigma_{Agm}(\tau) \rangle ^{\mu_2} \delta(c\tau+r_1') 
\nonumber\\&
+ \sqrt{c\gamma_{2_A}}  \langle \sigma_{Agm}(\tau) \rangle ^{\mu_1} \delta(c\tau+r_2') 
\nonumber\\&
- \left( i g_{M_{21}} +\frac{\sqrt{\gamma_{2_A}\gamma_{1_B}}}{2}  \right)  \langle \sigma_{Agm}(\tau) \sigma_{Bgm}(\tau)   \rangle ^{\mu_1 \mu_2}
\nonumber\\&
- \frac{ \sqrt{\Gamma_{3_A}\Gamma_{3_B}}}{2} \langle  \sigma_{Bge}(\tau)    \rangle ^{\mu_1 \mu_2},
\label{<sigma_1ge>^mu_1,25}
\end{align}
where we have used the relation of Eqs.~(\ref{a_r(t)}) and (\ref{b_r(t)}), and ignored the components having no contribution to $ \langle \sigma_{Age}(\tau)  \rangle ^{\mu_1\mu_2}$.
In the same manner, we can write down the equations of motion for other operators.
Using initial conditions, e.g., $ \langle \sigma_{Age}(0)  \rangle ^{\mu_1\mu_2}=0$, we have solved these simultaneous equations, and then determined the propagator by substituting the solutions into Eq.~(\ref{eq.uppropagator5}).
Then, the up-conversion probability $P_{\rm uc}$ and two-photon emission probability $P_{\rm two}$ are calculated as 
\begin{eqnarray}
P_{\rm uc}&=&\int dr |h(r;\tau)|^2,
\\
P_{\rm two}&=&\iint dr_1 dr_2 |g(r_1,r_2;\tau)|^2,
\end{eqnarray}
where they satisfy $P_{\rm uc}+P_{\rm two} = 1$.

\subsection{Input two-photon with correlation \label{subsec.input}}
\begin{figure}[t]
\includegraphics[width=80mm]{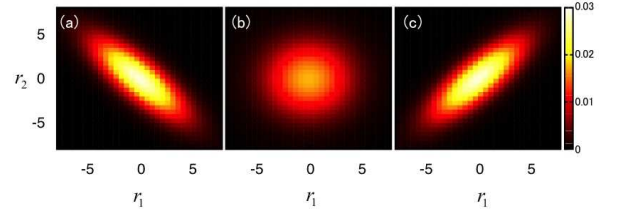}
\caption{Density plots of the input two-photon wave function of (a) spatially anti-correlated
($\rho$ = -0.8), (b) spatially uncorrelated ($\rho$ = 0), and (c) spatially correlated ($\rho$ = 0.8) states.}
\label{correlationphoton}
\end{figure}

It is known that correlated photons enhance two-photon process owing to allowing simultaneous excitations, thus utilizing of photons with quantum correlation for two-photon processes has attracted much attention~\cite{TPA1,TPA2,TPA3,TPA4,PhysRevLett.103.123602,TPA5}. Besides, because entangled photons are key issues in quantum information technology~\cite{NC}, the study of nonlinear responses by the correlated photons is not only interesting in itself, but also can contribute to further development in photoscience. Therefore, in order to discuss the effect of the photon correlation for the nonlinear process, we assume that the input two-photon wave function is expressed by bi-variable Gaussian pulse as
\begin{align}
f(r_{1},r_{2})
&=
\frac{\exp 
\left[ -\frac{\bar{r}_1^2+\bar{r}_2^2-2\rho \bar{r}_1 \bar{r}_2}{4(1-\rho^2)d^2} 
+i \frac{\omega_0}{c}(\bar{r}_1+\bar{r}_2)
\right] }
{(2 \pi )^{1/2}d(1-\rho^2)^{1/4}},
\end{align}
where $\bar{r}=r-a$ is the distance from the initial position $a$, and the parameter $\rho$ denotes the correlation between two photons. The pulse length $d$ corresponds to the temporal coherence length of the photon, and recent experiments reported that entangled photons or single photon sources generate photons with the long coherence length of $10^{-4} \sim 10^2$ m \cite{photoncoherence1,photoncoherence2,photoncoherence3,photoncoherence4,photoncoherence5,photoncoherence6}. Although the coherence length of sunlight is approximately several hundreds of nanometers \cite{sunlightcoherence1}, photons with that of the order of cm are obtained by the spectral filtering technique \cite{sunlightcoherence2}. The frequency of the pulse $\omega_0$ is set to the localized surface plasmon resonance. Figure \ref{correlationphoton} shows the density plots $|f(r_{1},r_{2})|^2$ for the wave function at (a) $\rho=-0.8$, (b) $\rho=0$, and (c) $\rho=0.8$. When the correlation parameter $\rho$ is equal to 0, the input two photons can be decoupled. On the other hand, as $\rho$ gets close to $1$ $(-1)$, they are in strongly correlated (anti-correlated) state. Such an entangled photon-pair can be actually generated using spontaneous parametric down-conversion~\cite{QD&I}. We discuss the dependence of up-conversion processes on the correlation of input two-photon in Sec.~\ref{subsec.input-correlation}. In other sections, we assume that input two-photon is non-correlated, i.e., $\rho=0$.

\section{enhanced up-conversion \label{sec.results}}
We numerically calculate the probability of the up-conversion in molecules coupled to an antenna.
Hereafter we basically assume that the frequencies of the plasmon mode and molecules are set to be $\omega_p=\omega_{m_A,m_B}=\omega_{e_A,e_B}/2$. 
Because the relaxation constant of the plasmon mode is large compared to the other rates, we use $\Gamma_{3_A,3_B}/\Gamma_p=0.2$, $\gamma_{1_A,1_B}/\Gamma_p=0.01$, $\gamma_{2_A,2_B}/\Gamma_p=0.001$. In addition, for simplicity, we assume that both the antenna-molecule coupling constants are equal, i.e., $g_{p_A}=g_{p_B}=g_p$. 
We set the pulse length to be $d\Gamma_p/c=7$, which corresponds to $d=434$ ${\rm \mu m}$ for $\hbar\Gamma_p=20$ ${\rm meV}$.

\subsection{Effect of inter-molecule correlation \label{subsec.inter-molecule}}
\begin{figure}[bht]
\includegraphics[width=80mm]{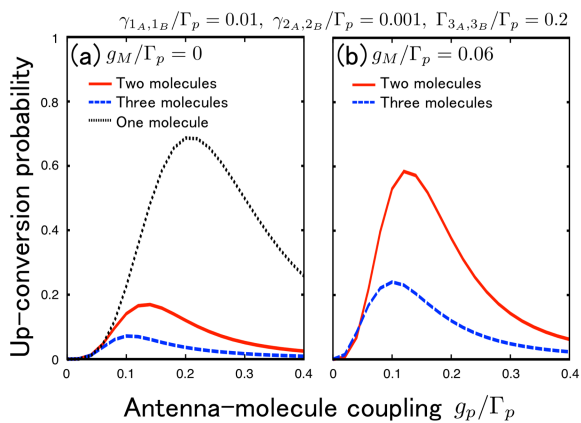}
\caption{Probability of up-conversion is plotted against the antenna-molecule coupling at the inter-molecule coupling constant (a) $g_M=0$ and (b) $g_M/\Gamma_p=0.06$. The solid, dashed, and doted lines express the probability in the case that the quantum complex system contains two molecules, three molecules and one molecule, respectively. We find that although the probability reduces with increasing the number of the molecules, the inter-molecule coupling enhances the probability. }
\label{multiUP}
\end{figure}
\begin{figure}[t]
\includegraphics[width=80mm]{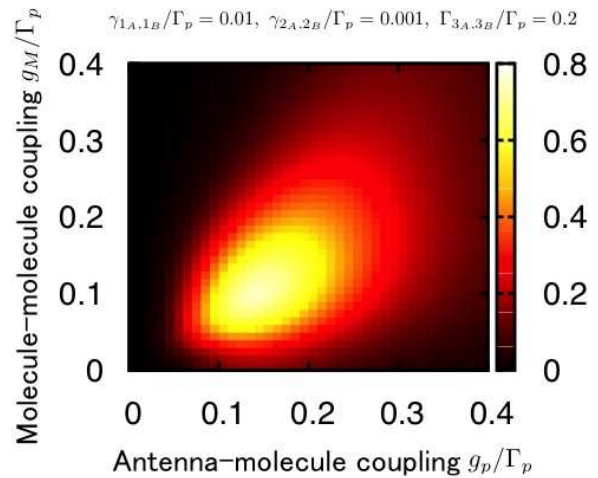}
\caption{Coupling dependence of the up-conversion probability on the two molecules, which are coupled to an antenna. The probability is plotted against the antenna-molecule coupling $g_p$ and the molecule-molecule coupling $g_M$. We find the optimal ranges for both the couplings. }
\label{2mUPcolor}
\end{figure}
\begin{figure}[!h]
\includegraphics[width=80mm]{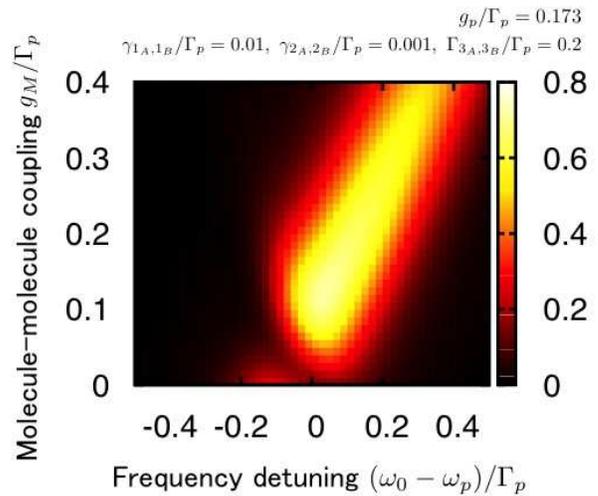}
\caption{Frequency detuning dependence of the up-conversion probability on the two molecules coupled to an antenna. The probability is shown as functions of the molecule-molecule coupling $g_M$ and the frequency detuning of the input photon from the resonances of the plasmon and the molecule. It is found that the optimal range of the coupling depends on the frequency detuning.}
\label{2mUPcolor2}
\end{figure}

\begin{figure}[!h]
\includegraphics[width=80mm]{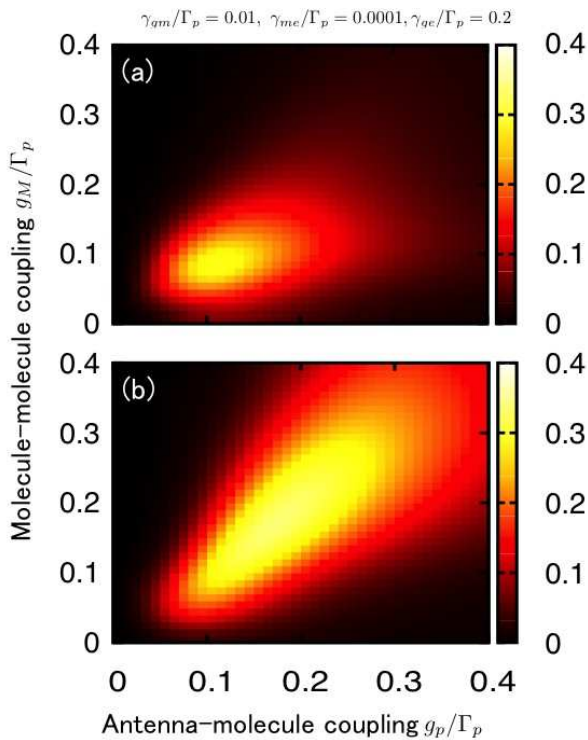}
\caption{Up-conversion probability is plotted against the antenna-molecule coupling $g_p$ and the molecule-molecule coupling $g_M$ in the case of containing (a) three molecules and (b) four molecules within the coupled system.}
\label{3m4mUPcolor}
\end{figure}

We investigate the effects of the presence of more than one molecule on the up-conversion process. The probability of the up-conversion for inter-molecule coupling $g_M=0$ is shown in Fig.~\ref{multiUP}(a).
In the case of containing one molecule with in the quantum complex system (black dotted line), we can see that the probability is enhanced at the optimal antenna-molecule coupling, where the loss of the antenna is suppressed by the quantum interference~\cite{OsakaPRL2014}. However, as the number of the molecules within the quantum system increases, the up-conversion probability decreases.
This is because the second and third molecules inhibit the process of two photons in a molecule. On the other hand, Figure~\ref{multiUP}(b) shows the probability of the up-conversion for inter-molecule coupling $g_M/\Gamma_p=0.06$. In comparison with Fig.~\ref{multiUP}(a), it is found that the inter-molecule coupling enhance the probability. In order to examine the coupling dependence of the up-conversion process, we show the probability as a function of the molecule-molecule coupling. At first, we consider the up-conversion in two molecules, which are coupled to an antenna. Figure \ref{2mUPcolor} shows the probability plotted against the antenna-molecule and molecule-molecule coupling, and reveals that the up-conversion is enhanced at both the optimal couplings. In Fig.~\ref{2mUPcolor2}, the probability is calculated as functions of the molecule-molecule coupling $g_M$ and the detuning of the input photon energy from the resonances of the plasmon and the molecule. This result states that the optimal range of the coupling depends on the frequency detuning. Figure \ref{3m4mUPcolor} shows the couplings dependence of the up-conversion probability in the case of containing more than two molecules within the coupled system. As is the case in Fig.~\ref{2mUPcolor}, it is confirmed that there are the optimal ranges of the both couplings. These results mean that controlling both the antenna-molecule and the inter-molecule couplings leads to high-efficient up-conversion even though the number of molecule is more than two.

\subsection{Essence of enhancement \label{subsec.essence}}
\begin{figure}[!h]
\includegraphics[width=80mm]{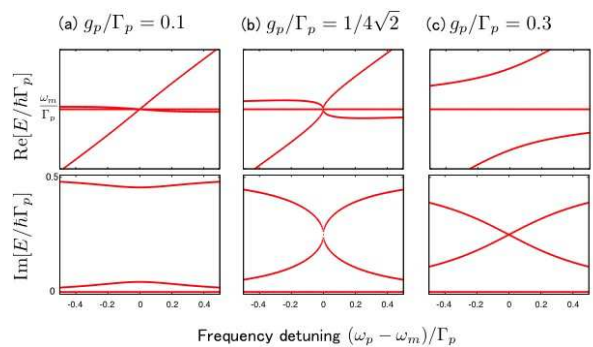}
\caption{Eigenvalues of the coupled modes for the antenna-molecule coupling constant (a) $g_p/\Gamma_p=0.1$, (b) $g_p/\Gamma_p=1/4\sqrt{2}$, and (c) $g_p/\Gamma_p=0.3$. Here the inter-molecule coupling is fixed to $g_M=0$. The horizontal axis is the detuning between the resonant frequency of the antenna mode and that of the first excited state of the molecule $|m_{A(B)}\rangle$. The upper (lower) three figures show the real (imaginary) part of the eigenmodes. This complex frequency of the dark mode is always constant because it does not include the antenna mode. In the upper figures, one can find an anti-crossing (Rabi splitting) when $g_p / \Gamma_p$ is larger than $1/4\sqrt{2}$. On the other hand, when $g_p / \Gamma_p$ is smaller than $1/4\sqrt{2}$, one can find a crossing. Then, the energies of upper and lower branches become equal at $\omega_p=\omega_m$. In the lower figures, one can find a crossing when $g_p / \Gamma_p$ is larger than $1/4\sqrt{2}$. Then, the radiative decay rate of upper and lower branches become equal at $\omega_p=\omega_m$. Therefore, the two bright modes oscillate in the same frequency and with the same time constant at $g_p / \Gamma_p=1/4\sqrt{2}$ and $\omega_p=\omega_m$. This is the condition for the quantum interference in the coupled system.}
\label{fig.multieigenA-M}
\end{figure}
We discuss the essence of the enhancement at the optimal inter-molecule coupling. Here, for simplified discussions, we ignore the radiative decay rates of the molecules, i.e., $\gamma_{1_A,1_B}=\gamma_{2_A,2_B}=\Gamma_{3_A,3_B}=0$. In addition, the resonance energies of the two molecules are equal to each other, i.e., $\omega_{m_A}=\omega_{m_B}=\omega_m,~\omega_{e_A}=\omega_{e_B}=\omega_e$.

The eigenmodes of the coupled system for the different antenna-molecule coupling constants $g_p$ is shown in Fig.~\ref{fig.multieigenA-M}, where the inter-molecule coupling is fixed to $g_M=0$. The upper (lower) three figures show the real (imaginary) part of eigenmodes, which corresponds to the frequency (decay rate) of the coupled modes. Here, this complex frequency of the dark mode is always constant because it does not include the antenna mode. The other two modes efficiently interfere constructively and destructively at the crossing point, which appears at the optimal antenna-molecule coupling. This is because the two modes oscillate in the same frequency and with the same time constant. Owing to this quantum interference in the coupled system, when the destructive interference occur in the antenna mode, it is possible to make only the molecular polarization oscillate. Then, the large loss in antenna is suppressed and we can achieve the efficient few-photon nonlinear responses~\cite{OsakaPRL2014}. However, in the case that more than one molecule is present, these conditions are not sufficient to induce the high-efficient up-conversion. The reason is explained as follows. Because input two-photon interacts with a number of molecules, the expected values of the population of the individual molecules decrease. Therefore, whereas photons efficiently excite a number of molecules owing to this quantum interference, strong photon-photon interaction in one molecule, which is of importance to nonlinear optical processes, is inhibited. Figure~\ref{multiUP}(a) corresponds to the above case.

\begin{figure}[t]
\includegraphics[width=80mm]{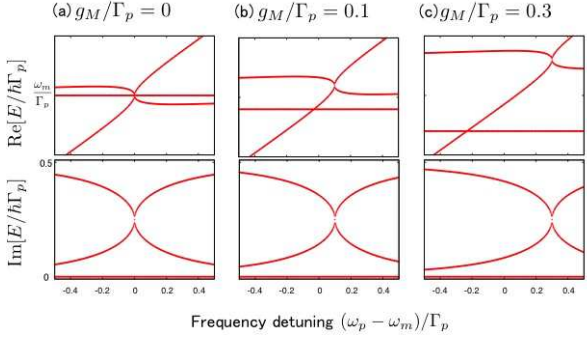}
\caption{ Eigenvalues of the coupled modes for the antenna-molecule coupling constant (a) $g_M/\Gamma_p=0$, (b) $g_M/\Gamma_p=0.1$, and (c) $g_M/\Gamma_p=0.3$. Here the antenna-molecule coupling is fixed to $g_p=1/4\sqrt{2}$. The horizontal axis is the detuning between the resonant frequency of the antenna and that of the first excited state of the molecule $|m_{A(B)}\rangle$. The upper (lower) three figures show the real (imaginary) part of the eigenmodes. In the upper figures, it is found that the crossing point shifts due to the splitting of the mode frequencies induced by the molecule-molecule coupling.
Accordingly, the offending interference from the other molecule is avoided.}
\label{fig.multieigenM-M}
\end{figure}
Figure \ref{fig.multieigenM-M} shows the real and imaginary parts of the eigenmodes for the different molecule-molecule coupling constants $g_M$, where antenna-molecule coupling is fixed to $g_p=1/4\sqrt{2}$.
From the upper and lower figures, one can see that the crossing point shifts with increasing the molecule-molecule coupling. This shift is due to the splitting of the molecular modes induced by the inter-molecule coupling. Actually, Figure \ref{2mUPcolor2} shows that the optimal energy of input photons for the up-conversion moves to the higher frequency side with increasing the inter-molecule coupling.
From these results, we interpret that the degeneracy of the molecular levels is lifted by the inter-molecule coupling so that the offending interference due to the presence of more than one molecule is avoided.
Therefore, by controlling both the antenna-molecule coupling and the molecule-molecule coupling, we can achieve the efficient few-photon nonlinear responses in the presence of a number of the molecule.

\subsection{Dependence on correlation of input photons \label{subsec.input-correlation}}
\begin{figure}[!h]
\includegraphics[width=80mm]{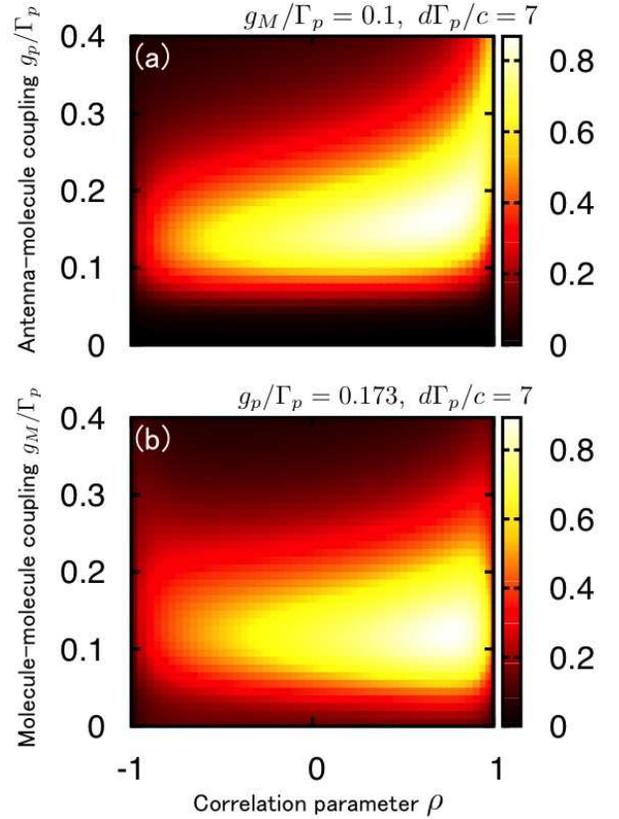}
\caption{\label{fig.rho-up} Dependence of the up-conversion on the input photon-correlation.  (a) The up-conversion probability is plotted as a function of input photon correlation and antenna-molecule coupling at $g_M/\Gamma_p=0.1$. (b) The probability is plotted as a function of input photon correlation and molecule-molecule coupling at $g_p/\Gamma_p=0.173$. When input two photons are correlated in space (i.e., $\rho$ is close to 1), the up-conversion process is facilitated.}
\end{figure}
We discuss the dependence of the up-conversion on correlations between input two photons. In Fig.~\ref{fig.rho-up}(a), the up-conversion probability is plotted as a function of the input two-photon correlation and the antenna-molecule coupling. In Fig.~\ref{fig.rho-up}(b), the probability is plotted as a function of the input two-photon correlation and the molecule-molecule coupling. These figures show that the up-conversion process is facilitated when input two photons are correlated in space, i.e., the correlation parameter $\rho$ is close to 1. This is because the spatial correlation allows two photons to interact the coupled system almost simultaneously and excite the molecules sequentially. The sequential excitation is of importance to the up-conversion process. Accordingly, the correlation of input photons spreads the optimal regime of the antenna-molecule and molecule-molecule coupling for the efficient up-conversion.

\section{conclusion \label{sec.conclusion}}
Larger antenna-molecule coupled system, which involves more than one molecule, is suitable in order to enlarge the total scattering probability between photons with micrometer-scale wavelength and molecules nearby nanometer-scale metallic antennas. However, the presence of multiple molecules seems to damage nonlinear optical processes because of the inhibition of photon-photon interaction in one molecule, when the system is irradiated by weak light, which contains only a few photons. Therefore, we have theoretically studied two-photon up-conversion in a quantum complex system where more than one molecule is coupled to a metallic nanoantenna. As a result, we have confirmed that the probability decreases with increasing the number of the molecules. However, we have shown that controlling the inter-molecule coupling, which is enhanced near the metallic nanoantenna, resolves the difficulty by lifting the degeneracy of the energy levels of the molecular system. Therefore the total design of the antenna-molecule and inter-molecular coupling enhances few-photon nonlinear responses in the large antenna-molecule coupled system. These results will open new avenues for nonlinear optical devices to realize single-photon control techniques, e.g., wavelength conversion of single photons and single photon switching.

\section*{Acknowledgements\label{sec.acknowledge}}
This work was partially supported by a Grant-in-Aid for JSPS Fellows No.~25$\cdot$09308 and for Challenging Exploratory Research No.~15K13505 from the Japan Society for
Promotion of Science (JSPS).


\providecommand{\noopsort}[1]{}\providecommand{\singleletter}[1]{#1}%

\end{document}